\documentclass[12pt]{article}
\usepackage{amsfonts}
\usepackage{amsmath}
\usepackage{graphicx}

\newtheorem{theorem}{Theorem}

\newtheorem{algorithm}[theorem]{Algorithm}

\newtheorem{definition}[theorem]{Definition}

\newtheorem{proposition}[theorem]{Proposition}
\newtheorem{remark}[theorem]{Remark}

\newenvironment{proof}[1][Proof]{\noindent\textbf{#1.} }{\ \rule{0.5em}{0.5em}}
\DeclareMathOperator*{\argminA}{arg\,min}
\DeclareMathOperator*{\argmaxA}{arg\,max}
\begin{document}

\title{A construction of a graphical model}
\author{Konrad Furma\'nczyk, Warsaw University of Life Sciences, \\konrad\_furmanczyk@sggw.edu.pl}

\maketitle

\begin{abstract}
	We present a nonparametric graphical model. Our model uses an undirected graph that represents conditional independence for general random variables defined by the conditional dependence coefficient (Azadkia and Chatterjee (2021)). The set of edges of the graph are defined as 
	$E=\{(i,j):R_{i,j}\neq 0\}$, where $R_{i,j}$ is the conditional dependence coefficient for $X_i$ and $X_j$ given $(X_1,\ldots,X_p) \backslash \{X_{i},X_{j}\}$.
	We propose a graph structure learning by two steps selection procedure: first, we compute the matrix of sample version of the conditional dependence coefficient  $\widehat{R_{i,j}}$; next, for some prespecificated threshold $\lambda>0$  we choose an edge $\{i,j\}$ if
	$ \left|\widehat{R_{i,j}} \right| \geq \lambda.$ The graph recovery structure has been evaluated on artificial and real datasets. We also applied a slight modification of our graph recovery procedure for learning partial correlation graphs for the elliptical distribution.
\end{abstract}


\section{An Introduction}
Graphical dependency models have been intensively studied in the two last decades (Lauritzen (1996), Drton and Perlman (2007), Maathuis et al. (2020)). The core of this research consists of undirected Gaussian graphical models based on a multivariate normal distribution, Bayesian networks, vine copulas (Bedford and Cooke (2002)) and recently undirected graphical models for nonparanormal distributions (Liu, Lafferty, and Wasserman (2009)). A graphical model is a convenient tool for the exploration of multivariate dependence. Most of those models explaining conditional independence. Only a few models, for example a covariance graph model (Cox and Wermuth (1996)) considers marginal independence restrictions. Several papers presents graphical models for elliptical distribution (Finegold and Drton (2009), Vogel and Fried (2011), Rossell and Zwiernik (2021)). In this setting the model is given by zeros in the inverse covariance (equivalently, by vanishing partial correlations). Outside Gaussian case, the corresponding partial correlation graphs (PG) cannot be interpreted in terms of conditional independence (Rossell and Zwiernik (2021)). 
In our work, we first investigate the conditional independence model (called the General Graphical Model (GGM)) by the coefficient of conditional dependence (Azadkia and Chatterjee (2021)). The GGM is a natural generalization of the graphical Gaussian model (Lauritzen (1996)). We learn an underlying graph structure using a two steps selection procedure which we present in Subsection 2.3. In the first step, we perform estimation of the matrix $\mathbf{R}$ which is a matrix of the conditional correlation coefficients. In the second step, for some chosen a threshold $\lambda$ we finally select the model $\widehat{E^{T}}=\{(i,j):|\widehat{R}_{i,j}| > \lambda\}$.
In the second step, similarly to Rothman et al. (2009) we may equivalently consider a sparse estimation of matrix $\mathbf{R}$ using $L^1$ penalty estimation (\ref{sigma}). 
Our main result is Theorem \ref{MainTh} which presents consistency model selection. In a Simulation Study and Real data analysis, we are checking the quality of the proposed selection procedure for the GGM in comparison with the Glasso and the nonparanormal graphical model. Finally, in Section 3 we present similar selection procedure for PG models and we formulate general theorem of consistency model selection-Theorem \ref{PG1}. All proofs are given in the Appendix. 
\\Proposed recovery method are very fast. The most time complexity part of our algorithm is computing the coefficient of conditional dependence (Azadkia and Chatterjee (2021)), which is fully nonparametric and can be estimated from data in time $O(n\log{n})$, where $n$ is the sample size. 
\\An alternative method of recovering the unknown graph structure from the data is to use a conditional independence test combined with some kind of multiple testing correction (for example, as in the case of Gaussian models used Drton and Perlman (2007)). However, following the result in Shah and Peters (2018), in the continuous case conditional independence testing is a hard problem, that means all existed tests have very small power (asymptotically not larger than a fixed significant level $\alpha$). A natural candidate for use is the Kernel Conditional Independence Test (Zhang et al. (2011)), which in addition of low power requires the use of a lot of computational cost with multidimensional data. A review of conditional independence tests can be found in Mielniczuk (2022).
Recently Baptista et al. (2021) invented SING algorithm which in non-Gaussian case estimates the joint probability density using a deterministic coupling, induced by a triangular transport map, and iteratively exploits sparse structure in the map to reveal sparsity in the graph. They showed that this algorithm recovers the graph structure even with a biased approximation to the joint density. They also characterize conditional independence, by a score based on integrated Hessian information from the joint log-density. However, their algorithm is numerically quite complex.

\section{GGM (General Graphical Model)}

In this section, we present the dependency structure of the GGM and we show the recovery of this structure by the conditional correlation coefficients (\ref{estymte}).

\subsection{Graph structure}
Let $G=(V,E)$ be an undirected graph, where $V=\{1,2,\ldots ,p\}$ - set of nodes, $E$ - a
set of undirected edges (pairs $\left( i,j\right) \in E$). Let $\mathbf{X}%
=(X_{1},X_{2},\ldots ,X_{p})$ be a random vector in $R^{p}$. We define the pairwise (Markov property) conditional independence structure of our graph as
\begin{equation*}
\left( i,j\right) \notin E\Longleftrightarrow X_{i}\perp X_{j}\mid \mathbf{%
	X}_{-{i,j}}, \text{ where } \mathbf{X}_{-{i,j}}%
=\mathbf{X\backslash \{}X_{i},X_{j}\mathbf{\}}.
\end{equation*}

When the density $\pi$ of $\mathbf{X}$ is strictly positive, i.e. $\pi(\mathbf{x})>0$ for all $\mathbf{x} \in R^p,$ the global and the pairwise Markov properties are equivalent (Lauritzen (1996)). The GGM generalizes the Gaussian graphical model (Lauritzen (1996)).

In our model, we use the coefficient of conditional dependence introduced by Azadkia and Chatterjee (2021). 
Let $\mu$ be the law of $X_i$. We use the following quantity as a measure of the degree of conditional dependence of $X_i$ and $X_j$ given $\mathbf{X}_{-{i,j}}$:\\
\begin{equation}\label{cond1}
T_{i,j|\mathbf{X}_{-{i,j}} }
= T(X_i,X_j|\mathbf{X}_{-{i,j}}) :=\frac{\int E(Var(P(X_i \geq t|X_j,\mathbf{X}_{-{i,j}})|\mathbf{X}_{-{i,j}}))d\mu(t)}{\int E(Var(1(X_i \geq t)|\mathbf{X}_{-{i,j}}))d\mu(t)}.
\end{equation}

The coefficient $T_{i,j|\mathbf{X}_{-{i,j}} }$ is undefined when the denominator equals zero. This happens if and only if $X_i$ is almost surely equal to a measurable function of $\mathbf{X}_{-{i,j}}$. We will ignore this case in our considerations. This coefficient has a natural interpretation as a nonlinear generalization of the partial $R^2$ statistic for measuring conditional dependence be regression. 

\medspace

We may observe that $0 \leq T_{i,j|\mathbf{X}_{-{i,j}}} \leq 1$ and in  Azadkia and Chatterjee (2021) the conditional independence is characterized as follows:\\

\begin{equation} \label{cond1a}
T_{i,j|\mathbf{X}_{-{i,j}}}=0 \Longleftrightarrow X_{i}\perp X_{j}\mid \mathbf{X}_{-{i,j}}.
\end{equation}

In the next section, we show recovery of  the GGM structure by an empirical version of the coefficient of conditional dependence.

\subsection{Recovery of graph structure}

 The true sparse model is given by $%
 E^{T}=\{(i,j):T_{i,j|\mathbf{X}_{-{i,j}}}%
 \neq 0\}$. 

\medskip

We consider the same estimator of $T_{i,j|\mathbf{X}_{-{i,j}}}$ as in Azadkia and Chattrjee (2021).
	We have data of $n$ i.i.d. copies $\mathbf{X}^{(1)}, \ldots, \mathbf{X}^{(n)}$ of the $p$-dimensional vector $(X_1,\ldots,X_p)$, where $n \geq 2$. For each $1\leq k \leq n$, let $N(k)$ be the index $l$ such that $\mathbf{X}^{(l)}_{-{i,j}}$ is the nearest neighbor of $\mathbf{X}^{(k)}_{-{i,j}}$ with respect to the Euclidean metric on $R^{p-2}$, where ties are broken uniformly at random. Let $M(k)$ be the index $l$ such that $\mathbf{X}^{(l)}_{-l}$ is the nearest neighbor of $\mathbf{X}^{(k)}_{-k}$ in $R^{p-1}$, again with ties broken uniformly at random. Let $R_k$ be the rank of $X_i^{(k)}$, that is, the number of $l$ such that $X_i^{(l)} \geq X_i^{(k)}$. If $p \geq 1$, then estimate of $T_{i,j|\mathbf{X}_{-{i,j}}}$ is

\begin{equation} \label{estymte}
T_{i,j|\mathbf{X}_{-{i,j}}}^{n}:=\frac{\sum_{k=1}^{n} (min(R_k,R_{M(k)}) - min(R_k,R_{N(k)}))}{\sum_{k=1}^{n}  (R_k-min(R_k,R_{N(k)}))}.
\end{equation}

To obtain a rate of convergence of $T_{i,j|\mathbf{X}_{-{i,j}}}^{n}$ to $T_{i,j|\mathbf{X}_{-{i,j}}}$, we need some assumptions about the distribution of random vector $\mathbf{X}$ (see Azadkia and Chattrjee (2021)).
\pagebreak 
 We consider the following assumptions.
\begin{description}
	\item[(A1)] There are non-negative real numbers $\beta$ and $C$ such that for any $t \in R, \\ \mathbf{x}, \mathbf{x'} \in R^{p-2}$ and $z, z' \in R$,
	\begin{multline}
		\\
		\sup_{1\leq i,j \leq p} \left| P(X_i \geq t|\mathbf{X}_{-{i,j}} = \mathbf{x}, X_j = z)-P(X_i \geq t|\mathbf{X}_{-{i,j}} =\mathbf{x'}, X_j = z')\right| \\
		\leq C(1 + \Vert \mathbf{x} \Vert^\beta+\Vert \mathbf{x'} \Vert^\beta  +\left|  z\right|^\beta+ \left| z'\right|^\beta)(\Vert \mathbf{x} - \mathbf{x'}\Vert^\beta + \left| z - z'\right|),
	\end{multline}
	and
	
	\begin{multline}
		\sup_{1\leq i,j \leq p} \left| P(X_i \geq t|\mathbf{X}_{-{i,j}} = \mathbf{x})-P(X_i \geq t|\mathbf{X}_{-{i,j}}  = \mathbf{x'})\right| 
		\\
		\leq C(1 + \Vert \mathbf{x} \Vert^\beta+\Vert \mathbf{x'} \Vert^\beta)(\Vert \mathbf{x} - \mathbf{x'}\Vert).
	\end{multline}
	
\end{description}

\begin{description}
	\item[(A2)] There are positive numbers $C_1$ and $C_2$ such that for any $t > 0$,
	\begin{equation*}
	\sup_{1\leq i,j \leq p} P(\Vert \mathbf{X}_{-{i,j}}\Vert \geq t) \text{, and }  \sup_{1\leq j\leq p}P(\left| X_j\right|  \geq t) \text{ are bounded by } C_1\exp(-C_2t).
	\end{equation*}
\end{description}
Conditions (A1)-(A2) are satisfied when $\mathbf{X}$ has normal distribution or when the support of $\mathbf{X}$ is a finite set. More general examples you may find in Azadkia and Chattrjee (2021), Proposition 4.2. 

\medskip

By a reasoning similar as in Azadkia and Chattrjee (2021), Theorem 4.1, we obtain
\begin{proposition}
 Under (A1)-(A2) we have
\begin{equation}\label{rzad}
	\sup_{1\leq i,j \leq p} \left| T_{i,j|\mathbf{X}_{-{i,j}}}^{n}- T_{i,j|\mathbf{X}_{-{i,j}}} \right| =O_P\left(\frac{(\log{n})^{p+\beta}}{n^{1/(p-1)}} \right),
\end{equation}
for some $\beta>0.$
 \end{proposition}
\subsection{Learning given sparse graph structure}

Now, we present our two steps selection procedure for graph structure. Our model selection procedure has two steps. In the first step, we perform estimation of the matrix $\mathbf{R}$ which is a matrix of the conditional correlation coefficients (\ref{cond1}).
 In the second step, for chosen threshold $\lambda$ we finally we select the model $\widehat{E^{T}}=\{(i,j):|\widehat{R}_{i,j}| > \lambda\}$.

\medskip

 In the second step, we may similarly to Rothman et al. (2009) consider $\widehat{\mathbf{\Sigma}}$ such as a sparse $L^1$ penalty estimation of matrix $\mathbf{R}$:
\begin{equation}\label{sigma}
\widehat{\mathbf{\Sigma}}=\argminA_ \mathbf{\Sigma} \frac{1}{2} \Vert \widehat{\mathbf{R}}-\mathbf{\Sigma} \Vert_F^2+\lambda\Vert \mathbf{\Sigma} \Vert_{1,off},
\end{equation} 
 where $\Vert \Vert_F$ is Frobenious norm, $\Vert \mathbf{\Sigma} \Vert_{1,off}=\sum_{i \neq j}\left| \Sigma_{ij}\right| $ and $\lambda>0$ is a tuning parameter. 
  
  \medskip

 It is known that we obtain a sparse solution of (\ref{sigma}) and 
  \medskip
  
 $
 \widehat{\Sigma}_{j,k} =\begin{cases}
 	sign(\widehat{R}_{j,k})max\left\lbrace \left|\widehat{R}_{j,k} \right|-\lambda,0 \right\rbrace & \text{for } j \neq k \\
 	\widehat{R}_{j,k} & \text{otherwise.}
 \end{cases}
$

For more technical details see Antoniadis and Fan (2001).
  \medskip
  
 Selected model in our graphical model is given by $ \widehat{E^{T}}=\{(i,j):\widehat{\Sigma}_{i,j} \neq 0\}$, where $\widehat{\Sigma}=(\widehat{\Sigma}_{i,j})$ is a sparse estimator of conditional correlation matrix $\mathbf{R}=(T_{i,j|\mathbf{X}_{-{i,j}}})$. Equivalently, we choose the following model $ \widehat{E^{T}}=\{(i,j):|\widehat{R}_{i,j}| > \lambda \}$.
 \\In real data application the tuning parameter $\lambda$ we may choose by cross validation or by calibration method.

\medskip

We consider an estimator $\widehat{\mathbf{R}}$ of the matrix $\mathbf{R}$ such that for all $k,r \in \{1,\ldots,p \} $ we have 
\begin{description}
	\item[($\Delta$)]  $\sup_{1\leq k,r \leq p} \left| \widehat{R}_{k,r}-R_{k,r} \right|  =O_P(a_n),$ for some $a_n$ as $n \rightarrow \infty$.
\end{description}	

Now we present some theoretical properties of our two-step selection procedure. 

\medskip
 
 \begin{proposition} \label{P1}
 If ($\Delta$) holds for some $a_n>0$ and $0 < \lambda \leq Ca_n$ for some constant $C>0$, then we have  
 \[ \sup_{1\leq k,r \leq p}  \left| \widehat{\Sigma}_{k,r}-R_{k,r} \right| = O_P(a_n).\]
 	
 \end{proposition}

  \begin{remark}
  If we use empirical conditional correlation coefficient (\ref{estymte}) for the GGM, then from (\ref{rzad}) we obtain condition ($\Delta$)  for $a_n=\frac{(\log{n})^{p+\beta}}{n^{1/(p-1)}}$ for some $\beta>0$. Azadkia and Chattrjee (2021) believe that for continuous variables  $a_n=\frac{1}{n^{1/(p-1)}}$ but they did not prove it.
  In the high dimensional case $p>n$, this error rate is very poor because $a_n \not \rightarrow 0 $ as $n \rightarrow \infty$. When $p$ is fixed and does not depend on $n$, then $a_n  \rightarrow 0 $ as $n \rightarrow \infty$ and the condition ($\Delta$) means that $a_n$ creates a rate of convergence of the empirical version of the conditional dependence coefficient (\ref{estymte}).
  \end{remark}

\medskip

Our main result is model selection consistency theorem. 

\begin{theorem} \label{MainTh} Suppose ($\Delta$) holds for some sequence $(a_n)$ such that: there exists a positive constant $C$, $Ca_n \leq \lambda \leq r_{min}-Ca_n$, where $r_{min}=min\{|R_{i,j}|: R_{i,j}\neq 0, i\neq j \}$ and $r_{min}>2Ca_n$. Then, 
	$$P(\widehat{E^T}=E^T) \rightarrow 1 $$ as $n \rightarrow \infty.$

\end{theorem}

\bigskip

\subsection{A Simulation study}

To check the quality of our selection procedure simulation study is carried out.

\medskip

We simulate random vectors $(X_1^i,\ldots, X_p^i)$, where $p=100$, $i \in \left\lbrace  1,\dots,n\right\rbrace $ for $n=100,200$, where we generate independently $n$ observations $X_j^i$ from the standard normal distribution $N(0,1)$ for $j \geq 7$ in models (M1)-(M4) and for  $j \geq 13$ in models (M5)-(M6) which create the following graphical models:
\begin{itemize}
	\item[M1]  We generate independently $X_2^i$ from the exponential Ex(1) distribution, $X_4^i$ from the t-Student distribution with 3 df and  $X_6^i$ from the exponential Ex(3) distribution, and $X_3^i=0.1\cdot X_4^i+\epsilon_1^i$,  $X_1^i=0.2\cdot X_2^i+X_3^i+\epsilon_2^i$,  $X_5^i=0.1\cdot X_6^i+\epsilon_3^i$, where $\epsilon_1^i,\epsilon_2^i,\epsilon_3^i$ are generated independently from $N(0,1)$ for  $i \in \left\lbrace  1,\dots,n\right\rbrace $;
	\item[M2]  We generate independently $X_2^i, X_4^i,X_6^i$ from $N(0,1)$, and $X_3^i=0.1\cdot X_4^i+\epsilon_1^i$,  $X_1^i=0.2\cdot X_2^i+X_3^i+\epsilon_2^i$,  $X_5^i=0.1\cdot X_6^i+\epsilon_3^i$, where $\epsilon_1^i,\epsilon_2^i,\epsilon_3^i$ are generated independently from $N(0,1)$ for  $i \in \left\lbrace  1,\dots,n\right\rbrace $;
	\item[M3]  We generate independently $X_2^i$ from the exponential Ex(1) distribution, $X_4^i$ from the t-Student distribution with 3 df and  $X_6^i$ from the exponential Ex(3) distribution, and $X_3^i=0.1\cdot exp(X_4^i)+\epsilon_1^i$,  $X_1^i=0.2\cdot sin(X_2^i)+sin(X_3^i)+\epsilon_2^i$,  $X_5^i=0.2\cdot exp(X_6^i)+\epsilon_3^i$, where $\epsilon_1^i,\epsilon_2^i,\epsilon_3^i$ are generated independently from $N(0,1)$ for  $i \in \left\lbrace  1,\dots,n\right\rbrace $;
	\item[M4] We generate independently $X_2^i, X_4^i,X_6^i$ from $N(0,1)$, and $X_3^i=0.1\cdot exp(X_4^i)+\epsilon_1^i$,  $X_1^i=0.2\cdot sin(X_2^i)+sin(X_3^i)+\epsilon_2^i$,  $X_5^i=0.2\cdot exp(X_6^i)+\epsilon_3^i$, where $\epsilon_1^i,\epsilon_2^i,\epsilon_3^i$ are generated independently from $N(0,1)$ for  $i \in \left\lbrace  1,\dots,n\right\rbrace $.
	\item[M5] We generate independently $X_1^i$ from Ex(1) and $X_2^i$ from Ex(3), and for $1\leq j\leq 10, X_{j+2}^i=0.2\cdot X_j^i+0.3\cdot X_{j+1}^i+\epsilon_j^i$, where $\epsilon_j^i$ are generated independently from $N(0,1)$ for  $i \in \left\lbrace  1,\dots,10\right\rbrace $.
	\item[M6] We generate independently $X_1^i, X_2^i$ from $N(0,1)$, and for $1\leq j\leq 10, X_{j+2}^i=0.2\cdot sin(X_j^i)+0.3\cdot sin(X_{j+1}^i)+\epsilon_j^i$, where $\epsilon_j^i$ are generated independently from $N(0,1)$ for  $i \in \left\lbrace  1,\dots,10\right\rbrace $.
\end{itemize}

In (M1)-(M2), we consider linear dependence models. In (M3)-(M4), we consider nonlinear dependence models. Only (M2) is Gaussian graphical model. Models (M1), (M3) are not satisfy (A1)-(A2), because we deal with high-tailed t-Student distribution. However, to test the robustness of our algorithm, we took these models into consideration. Fragments of the graphs of our models we may find on the Figure 1. In the left panel we have graph for models (M1)-(M4) and in the right panel we have graph for models (M5)-(M6). The graphs (M1)-(M4) have 4 cliques with 2 vertices and 94 cliques with 1 vertex, the graphs (M5)-(M6) have 6 cliques with 3 vertices, 2 cliques with 2 vertices and  90 cliques with 1 vertex.

\begin{figure}
	\includegraphics[width=11cm, height=7cm]{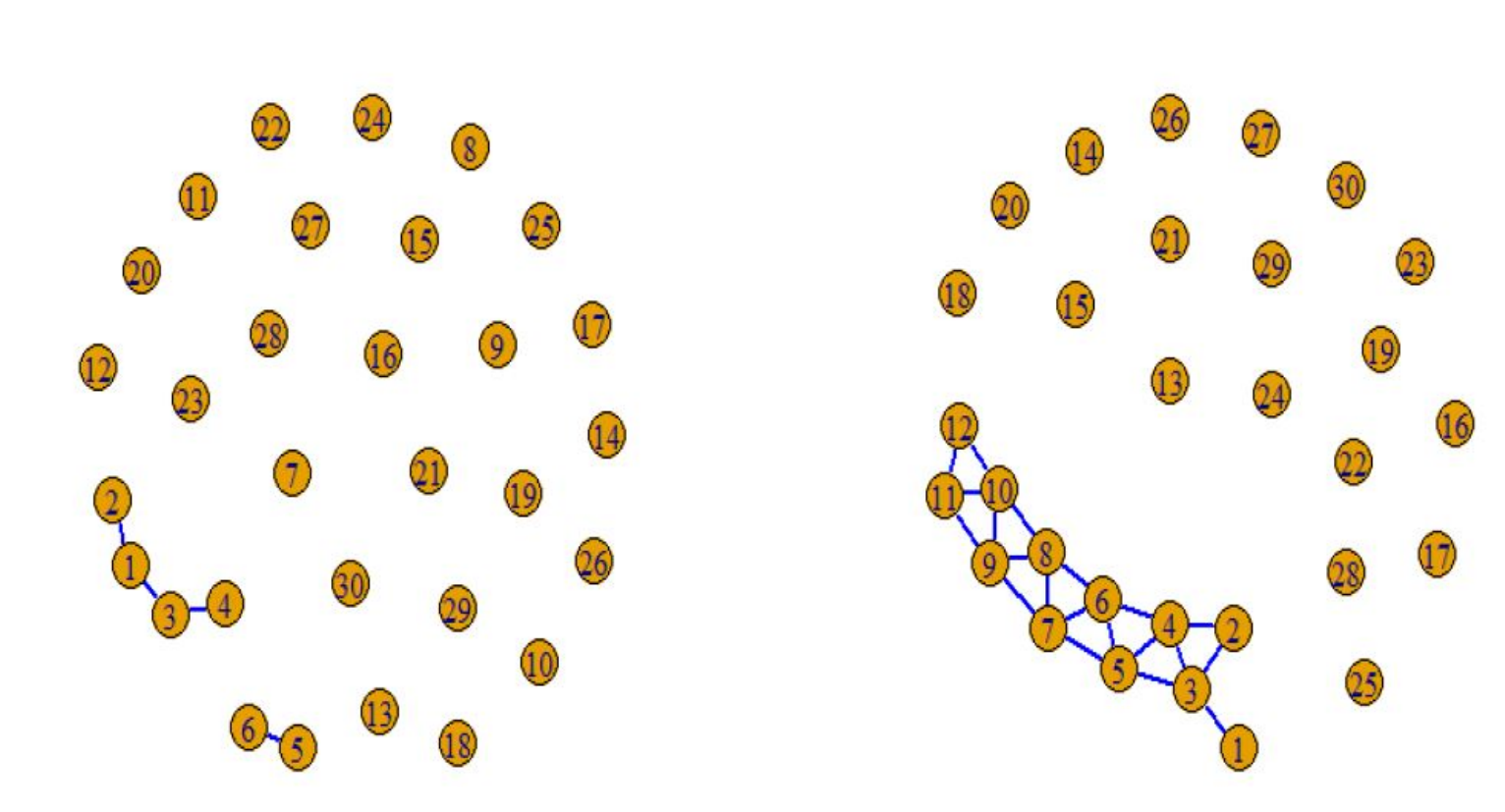}
	\caption{\label{Fig1} Fragments of the Graphs}
\end{figure}

\pagebreak

Let $\mathbf{R}=(R_{i,j})$ be a conditional correlation matrix in the GGM. From our constructions, we have that in models (M1)-(M4) we have $R_{1,2}\neq 0, R_{2,1}\neq 0, R_{1,3}\neq 0, R_{3,1}\neq 0, R_{3,4}\neq 0, R_{4,3}\neq 0, R_{5,6}\neq 0, R_{6,5}\neq 0, \text{and } R_{i,i}\neq 0 \text{ for } i \in \left\lbrace  1,\dots,p\right\rbrace $ and $R_{ij}$ equals zero otherwise. But in models (M5)-(M6) we have that $R_{1,2}\neq 0, R_{2,1}\neq 0, R_{j,j+2}\neq 0, R_{j+2,j}\neq 0, R_{j+1,j+2}\neq 0, R_{j+2,j+1}\neq 0  $ for $j \in \{1,\ldots,10 \} $ and $R_{ij}$ equals zero otherwise.
 \\Next, we perform our two step selection procedure. In more details, we compute estimator of the matrix  $\mathbf{R}$ using codec function from 'FOCI' package in R software, next the tuning parameter $\lambda$  in (\ref{sigma}) was chosen as $\lambda=1/n$. This tuning parameter we selected by calibration. This choice of the $\lambda$ parameter corresponds to the situation when $a_n=\frac{1}{n^{1/(p-1)}}$ (see Remark 2)). This guarantees a consistency model selection because the assumptions of Theorem 3 are satisfied.
  For comparision we apply the Glasso method and the Glasso for nonparanormal graphical model (see 'huge' R package and skeptic method for nonparanormal models (npn)). The nonparanormal extends Gaussian graphical models to semiparametric Gaussian copula models. Motivated by sparse additive models, the nonparanormal method estimates the Gaussian copula by marginally transforming the variables using smooth functions.

\medskip

In the GGM procedure first we compute
\begin{equation} \label{sigmax}
\widehat{\mathbf{\Sigma}_{i,j}}=sign(\widehat{R_{i,j}})(\left|\widehat{R_{i,j}} \right|-\lambda)_{+},
\end{equation}
 where $(a)_{+}=max\left\lbrace a,0 \right\rbrace $. Since the coefficient of conditional dependence (\ref{estymte}) is not symmetric, then we take a symmetric version of the matrix $\left|\widehat{R_{i,j}} \right|$ in our procedure as follows $$\left|\widehat{R_{i,j}} \right|:=max\{\left|\widehat{R_{i,j}} \right|,\left|\widehat{R_{j,i}} \right|\} $$ for all $i \ne j.$ \\It follows from (\ref{sigmax}) that our selection procedure chooses an edge $\{i,j\}$ if $$max\{\left|\widehat{R_{i,j}} \right|,\left|\widehat{R_{j,i}} \right|\} \geq \lambda .$$
 To summarise it, we use
 \begin{algorithm}
  a) first, we compute the matrix of sample version of the conditional dependence coefficient  $\widehat{R_{i,j}}$;
 \\ b) next, for some prespecificated threshold $\lambda>0$  we choose an edge $\{i,j\}$ if
 $$ max \left\{ \left|\widehat{R_{i,j}} \right|,\left|\widehat{R_{j,i}} \right| \right\} \geq \lambda.$$

\end{algorithm}
 
 To evaluate our model selection procedure and the comparative methods, we compute the True Proportion Rate (TPR) and the False Proportion Rate (FPR), where $$TPR=\frac{\#\left\lbrace (i,j): i \neq j,  \widehat{\mathbf{\Sigma}_{i,j}} \neq 0, R_{i,j} \neq 0 \right\rbrace }{\# \left\lbrace (i,j): i \neq j, R_{i,j} \neq 0 \right\rbrace }$$ and $$FPR=\frac{\#\left\lbrace (i,j): i \neq j, \widehat{\mathbf{\Sigma}_{i,j}} \neq 0, R_{i,j}=0 \right\rbrace }{\# \left\lbrace (i,j): i \neq j, R_{i,j}=0 \right\rbrace }.$$ The results from 100 replications of our models we present in Tables 1-2. For the Glasso method we use the smallest $\lambda$ parameter for the given path od 10 values. This choice for tuning parameter was the best for model selection.

\begin{table}[ht]
	\caption{The average values of the TPR from 100 replications} 
	\centering 
	\begin{tabular}{c c c c c c c} 
		\hline\hline 
		sample size & M1 & M2 & M3 & M4 & M5 & M6 \\ [0.5ex] 
		\hline 
	$n$=100 Glasso & 0.740 & 0.740 & 0.718 & 0.718 & 0.897 & 0.836 \\
	$n$=200 Glasso & 0.770 & 0.768 & 0.683  & 0.683 & 0.949 & 0.931 \\ 
	$n$=100 Glasso+npn & 0.693 & 0.718 & 0.733 &  0.733 & 0.881 & 0.843 \\
	$n$=200 Glasso+npn & 0.728 & 0.758 & 0.738  & 0.738 & 0.934 & 0.931 \\ 
	$n$=100 GGM & 0.978 & 0.980 & 0.960 & 0.973 & 0.968 & 0.976 \\
	$n$=200 GGM & 0.988 & 0.993 & 0.988  & 0.980 & 0.987 & 0.988 \\ [1ex]
		\hline 
	\end{tabular}
	\label{table:a} 
\end{table}

\begin{table}[ht]
	\caption{The average values of  the FPR from 100 replications} 
	\centering 
		\begin{tabular}{c c c c c c c} 
			\hline\hline 
			sample size & M1 & M2 & M3 & M4  & M5 & M6 \\ [0.5ex] 
			\hline 
			$n$=100 Glasso & 0.150 & 0.152 & 0.134 & 0.134 & 0.044 & 0.060 \\
			$n$=200 Glasso & 0.160 & 0.162 & 0.164  & 0.164 & 0.027 & 0.023 \\ 
			$n$=100 Glasso+npn & 0.173 & 0.161 & 0.124 &  0.124 & 0.050 & 0.058  \\
			$n$=200 Glasso+npn & 0.183 & 0.165 & 0.143  & 0.143 & 0.030 & 0.023 \\ 
			$n$=100 GGM & 0.000 & 0.000 & 0.000 & 0.000 & 0.003 & 0.003\\
			$n$=200 GGM & 0.001 & 0.001 & 0.001  & 0.000 & 0.004 & 0.004 \\ [1ex]
				\hline 
			\end{tabular}
	\label{table:b} 
\end{table}

In all models, we obtain that the FPR for the GGM method ranges from 0.000 to 0.004 and the TPR is between 0.95-0.993. The TPR for all models for the GGM methods is significantly greater than for the glasso and the nonparanormal graphical selection methods for models (M1)-(M6). Similarly, the lowest the FPR values and much lower than other methods were obtained for the GGM procedure. In all models and methods the TPR  increases with increasing the sample size. 

\subsection{Real data analysis}
We apply our considered method to gene expression data which contains 403 genes for 30 human brain samples) from the microarray study of Lu et al. (2004). 

We take gene expression data: lu2004 from 'care' R package, which consists of 403 genes for 30 samples from the microarray study of Lu et al. (2004). This data set contains measurements of the gene expression of 403 genes from 30 human brain samples.

\medskip

For lu2004 dataset where each gen is a vertex of our graph we obtained the following graph structures:

\medskip

\begin{itemize}
	\item  for the GGM ($\lambda=1/n$) -the graph with 471 edges
	\item  for the Glasso -the graph with 7517 edges
	\item  for the Glasso+npn -the graph with 7785 edges
	
\end{itemize}

\medskip

We may observe that the GGM method produce more sparse graph than comparative methods and we believe this method can be able to find true associations in large datasets.
	
\section{{Recovery of PG model for the elliptical distribution}}	

We introduce the basic definitions and notation. 

\begin{definition}
	A random vector $\mathbf{X} \in R^p$ has an elliptical distribution $E(\mu,\Sigma)$ if there exists $\mu \in R^p$ and positive semi-definite matrix $\Sigma$ such that the characteristic function of  $\mathbf{X}$ is of the form $t \rightarrow \phi(t^T\Sigma t)exp(i\mu^t t)$ for some $\phi:[0,\infty) \rightarrow R$.
\end{definition}	
	
It is not possible to define conditional independence in the elliptical family outside of the Gaussian case (see Proposition 2.5 in Rossell and Zwiernik (2021) and Theorem \ref{B}). 
\begin{theorem}\label{B}
	Theorem 3 (Baba et al. (2004)). Suppose $\mathbf{X} \sim E(\mu,\Sigma)$ and $X_i \perp X_j | X_C$ for some $i,j \in V$ and $C \subseteq \{1,\ldots,p\} \backslash \{i,j\} $. Then $\mathbf{X}$ is Gaussian.
\end{theorem}	
The partial correlation graph is defined as follows.
	
\begin{definition}
	The partial correlation graph (PG) is the graph $G=G(V)$ over vertex set $V=\{1,\ldots,p\}$	with an edge between $i \neq j$ if and only if $K_{i,j}\neq 0$.
\end{definition}

Next, we present characterization result of PG models. 

\textbf{Proposition} (Rossell and Zwiernik (2021))  $\mathbf{X} \sim E(\mu,\Sigma)$. Then $K_{i,j} =0 \Longleftrightarrow Cov(X_i,X_j|\mathbf{X}_{-{i,j}})=0$.  	

\medskip

Selected graph is given by $ \widehat{E^{T}}=\{(i,j):|\widehat{K}_{i,j}| \geq t_n \}$, where $\widehat{K}=(\widehat{K}_{i,j})$ is an estimator of the precision matrix $K$ and $t_n$ is a threshold. Now, we formulate our consistency model selection result. 

\begin{theorem}\label{PG1}
	Suppose $\Vert \widehat{K}-K\Vert_{\infty} =O_P(a_n) $ for some $a_n \rightarrow 0$ as $n \rightarrow \infty$ and the threshold  $t_n$ is such that $Ca_n \leq t_n \leq k_{min}-Ca_n$, where $k_{min}=min\{|K_{i,j}|: K_{i,j}\neq 0, i\neq j \}$ and $k_{min}>2Ca_n$ for some constant $C>0$. Then, 
	$$P(\widehat{E^T}=E^T) \rightarrow 1 $$ as $n \rightarrow \infty.$
\end{theorem}
In this selection procedure we need a good estimator of the Precision Matrix $K$. 
The Glasso framework still provides a consistent estimator for non-Gaussian data. 	
$$\widehat{K_{Glasso}}=\argmaxA_{K} \{logdetK-tr(SK)-\lambda \Vert K \Vert_{1,off} \}, $$ where $S$ sample covariance matrix. Under some regular conditions Ravikumar et al. (2011) showed that with high probability
$$\Vert \widehat{K}-K\Vert_{\infty}=O\left( \sqrt{\frac{\log p}{n}}\right)  $$ for sub-Gaussian case and
$$\Vert \widehat{K}-K\Vert_{\infty}=O\left(\sqrt{\frac{p^{\tau/m}}{n}}\right)  $$ for  $4mth$ moments bounded r.v. case and some $\tau>0$.	

\medskip

The same rate of convergence obtained Zhang and Zou (2014) for the DT estimator
$$\widehat{K_{DT}}=\argminA_{K\geq \epsilon I} \{1/2tr(K^2S)-tr(SK)+\lambda \Vert K \Vert_{1,off} \}. $$

\section{Conclusion}

The paper presents a concept of a graphical model. The GGM concerned the generalization of conditional independence (CI). A two-steps model selection procedure is also given. We can applied our nonparametric procedure GGM for conditional independence in the first step in the PC algorithm (Bayesian Network, see Spirtes et al. (2000)), where it learns from data a skeleton (undirected) graph. 
Based on the simulation, the proposed selection procedure seems to be reasonable. In all cases of the GGM method wins significantly with the Glasso method and nonparanormal model with the use of the skeptics method. Finally, we applied all considered methods for recovery graph structure from real dataset: gen expression data. In high dimensional gene expression dataset the GGM creates more sparse graph structure than the Glasso and similar graph structure as the nonparanormal model with the use of the skeptics method.

\section{Appendix}

\subsection{Proof of Proposition \ref{P1}}

 \begin{proof}
	From the KKT conditions, we have for $k \neq r$
	\[-\widehat{R}_{k,r}+\widehat{\Sigma}_{k,r}+\lambda sign(\widehat{\Sigma}_{k,r})=0  \] for $\widehat{\Sigma}_{k,r}\neq 0$ and 
	\[\left|\widehat{R}_{k,r}-\widehat{\Sigma}_{k,r} \right|\leq \lambda \] for $\widehat{\Sigma}_{k,r}=0$. 
	Therefore and from triangle inequality, we have 
	\[\left|\widehat{\Sigma}_{k,r}-R_{k,r} \right| \leq \left|\widehat{\Sigma}_{k,r}-\widehat{R}_{k,r} \right|+\left|\widehat{R}_{k,r}-R_{k,r} \right|,\] and 
	\[\left|\widehat{\Sigma}_{k,r}-R_{k,r} \right| \leq \lambda+\left|\widehat{R}_{k,r}-R_{k,r} \right|.\]
	Using  ($\Delta$) and our assumptions, we obtain Proposition \ref{P1} .
	
\end{proof} 
	
\subsection{Proof of Theorem \ref{MainTh}}

\begin{proof}
	First, we assume that $R_{i,j}=0$ for $i \neq j$. By ($\Delta$) and the assumptions of Theorem \ref{MainTh} we have that we probability tending to 1 with $n$ tends to $\infty$:
	$$| \widehat{R_{i,j}} |=| \widehat{R_{i,j}}-R_{i,j}|\leq \sup_{1\leq i,j \leq p}| \widehat{R_{i,j}}-R_{i,j}| \leq Ca_n \leq \lambda. $$ Then, with probability tending to 1, $\widehat{\Sigma_{i,j}}=0.$ 
	\\Now, we assume that $R_{i,j}\neq 0$ for $i \neq j.$ 
	Then, with probability tending to 1,  $$| \widehat{R_{i,j}} |\geq| R_{i,j}|-| \widehat{R_{i,j}}-R_{i,j}|\geq | R_{i,j}|-\sup_{1\leq i,j \leq p}| \widehat{R_{i,j}}-R_{i,j}|\geq r_{min}-Ca_n \geq \lambda. $$ Therefore with probability tending to 1, $\widehat{\Sigma_{i,j}}\neq 0.$ 'For those reasons, we obtain our thesis.
	
\end{proof}

\subsection{Proof of Theorem \ref{PG1}}
The proof is almost identical as the proof of Theorem \ref{MainTh} and we omit it.

\end{document}